\documentclass{article}
\usepackage{amsmath}
\usepackage{graphicx}

\newcommand{\dsum}{\sum}   
\newcommand{\dint}{\int}

\begin{document}

\title{S-wave and P-wave non-strange baryons in the \ potential model of QCD.%
}
\author{I.K.Bensafa$^{1,2}$, F.Iddir$^{1}$ and  L.Semlala$^{1}$ \\
$^{1}$ Laboratoire de Physique Th\'{e}orique \\
Universit\'e d'Oran Es-senia, 31100, Oran, Algeria \\
$^2$ Laboratoire de Physique Corpusculaire \\
Universit\'{e} Blaise Pascal/CNRS-IN2P3, F-63177 Aubi\`{e}re, France}
\maketitle

\begin{abstract}
In this paper, we study the nucleon energy spectrum in the ground-state
(with orbital momentum $L=0$) and the first excited state ($L=1$). The aim
of this study is to find the mass and mixing angles of excited nucleons
using a potential model describing QCD. This potential is of the
\textquotedblleft Coulombian+ linear\textquotedblright\ type and we take
into account some relativistic effects, namely we use essentially a
relativistic kinematic necessary for studying light flavors. By this model,
we found the proton and $\Delta (1232)$ masses respectively equal to ($968MeV
$, $1168MeV$), and the masses of the excited states are between $1564MeV$
and $1607MeV$. \newline
\end{abstract}

\section{Introduction}

In the quark model, the baryons are represented as bound states of three
quarks, confined by strong forces. Quantum Chromodynamics (QCD) aknowledged
as the theory of strong interactions, is however not able to describe
unambigously the strong-coupling regime, which requests alternative
theories, like Flux-Tube model, Bag model, QCD string model, Lattice QCD, or
phenomenological potential models. We will use in this work a
phenomenological potential model  describing explicitely the QCD
characteristics, and taking into account some relativistic effects. 

The potential model is essentially motivated by the experiment, and its wave
functions are used to represent the states of strong interaction and to
describe the hadrons. The most used is the harmonic oscillator potential
which requires very simple calculations and qualitatively in agreement with
the experimental data; on the other hand, the most usual potential models
are using non relativistic kinematics, which is convenient for the heavy
flavors systems, but cannot be suitable for the hadrons containing light
flavors.\newline

In this paper, we study the non-strange baryons spectrum in two different orbital momentum
states within a quark model framework. We calculate the mass of baryons in
the ground-state ($L=0$) with positive parity, and in the first excited
state ($L=1$) with negative parity. The method used is variational calculation 
using Schr\"{o}dinger equation. The baryon being a three-quarks system, we expose a
three-body problem, and the method used to resolve it. \newline

In section 2, we first introduce the Hamiltonian of our system. We use a
potential of the \textquotedblleft Coulombian+linear\textquotedblright\
form, which reproduces well the confinement and asymptotic freedom in QCD.
For the kinetic part of our Hamiltonian, we have added a semi-relativistic
correction. \newline

In section 3, we develop the wave function of our system corresponding to
ground-state using Jacobi variables. Then we diagonalize the Hamiltonian
matrix and minimise it to extract the baryon mass value. To improve this
value, we add the hyperfine correction which contains two terms. The first
one (called the contact term) takes into account the spin-spin interaction,
and it correlates the splitting of $\Delta -N$, $\Sigma -\Lambda $ in the
ground state. The second one, called the tensor term, averages to zero for
orbital momentum $L=0$.\newline

In section 4, we study the case of P-wave baryons with negative parity. We
use the same development as for the S-wave baryons to extract masses for the
first excited state $L=1$. In this case, the tensor term is operative, and
allows us to obtain mixing angles. Our results are then compared with the
work of Isgur and Karl \cite{isgur2}. \newline

\section{Potential model of QCD \ }

Several potential models were carried out until now for the determination of
the baryon mass spectrum in non-perturbative QCD. In general, one uses the
harmonic oscillator (\cite{isgur2}, \cite{isgur4}) approach in the quark
model . The result obtained is in agreement with the experimental data. This
approach provides a good determination of ground state and excited states of
the baryon energy spectrum. The Hamiltonian used is in the form:

\bigskip

\begin{equation}
H_{HO}=\dsum_{i<j=1}^{3}(E_{i}+V_{HO}(r_{ij})+H_{hyp})  \label{(HHO)}
\end{equation}

Where \textbf{$E_i$} is the kinetic part, and the harmonic oscillator
potential is in the following form :

\begin{equation}
V_{HO}(r_{ij})=\frac{1}{2}K\left\vert \overrightarrow{r_{i}}-\overrightarrow{%
r_{j}}\right\vert^2  \label{(VHO)}
\end{equation}

$H_{hyp}$ represents the hyperfine correction.\newline

A more appropriate potential is to be the so-called "Coulombic+ linear" one:
complicated non-perturbative effects are assumed to be largely absorbed into
the constituent quark masses and into a Lorentz scalar linear confinement
potential, the known short range behaviour of QCD is included in the one
gluon exchange "Coulombic" potential. In this approach the potential is
defined as follows \cite{isgur3}:

\begin{equation}
V(r_{ij})=-(-\frac{\alpha _{s}}{r_{ij}}+\frac{3}{4}\sigma r_{ij}+\frac{3}{4}%
c)F_{i}.F_{j}  \label{(Vrij)}
\end{equation}

The $F_{i}.F_{j}$ factor represents the color term, with $\alpha _{ij}=$ $%
\left\langle F_{i}.F_{j}\right\rangle =-\frac{2}{3}$.

The constants ($\alpha _{s}$, $\sigma $, c) are the phenomenological
parameters determined from an experimental fit.

In the present work, the calculation of the non-strange baryons spectrum is
carried out, by using the parameters of reference \cite{swanson}. The
results obtained will be discussed in sections 3 and 4.

\subsection{Variational Calculation}

The kinetic energy of the relativistic Schr\"{o}dinger equation describing a
system with three quarks is in the form :

\begin{equation}
E_{i}=\sqrt{P_{i}^{2}+m_{i}^{2}}  \label{(Ei)}
\end{equation}

But another term, essentially relativistic and more convenient for many-body
problems than the kinetic part of the Hamiltonian given by equation (\ref%
{(Ei)}), can be used: 
\begin{equation*}
(\frac{P_{i}^{2}}{2M_{i}}+\frac{M_{i}}{2}+\frac{m_{i}^{2}}{2M_{i}})
\end{equation*}

with the conditions:

\begin{equation*}
\frac{\partial E}{\partial M_{i}}=0
\end{equation*}
as in reference \cite{jacczko}, \cite{simonov}. \textbf{$M_{i}$} represents
the quark dynamical masses. They are heavier than the quark constituent
masses \textbf{$m_{i}$}.\newline

The Hamiltonian of the system is thus written (\cite{smlath}, \cite{smlala})
:

\begin{equation}
H=\dsum_{i<j=1}^{3}\left[ (\frac{P_{i}^{2}}{2M_{i}}+\frac{M_{i}}{2}+\frac{%
m_{i}^{2}}{2M_{i}})+V(r_{ij})\right]  \label{(H)}
\end{equation}

The calculation of the total energy of the system is carried out by the
variational method. This energy will be minimized w.r.t the variational
parameters of the test wave function and ($M_{i}$).\newline
The potential $V(r_{ij})$ is treated overall here like a non-perturbative
term. In reference \cite{capstick}, $V(r_{ij})$ is written in the form of
harmonic oscillator potential ($Kr_{ij}^{2}/2$) plus an unknown U-term
(treated like a perturbation) added to shift the energies of some states
(using the oscillator harmonic wave function).

In our work, we concentrate on the determination of the ground-state nucleon
mass, by minimizing the total energy. This method allows us to fix the
auxiliary dynamical masses (\textbf{$M_{i}$}) and the wave function
parameters. We use the same method to determine the excited-state nucleon
mass. The wave function parameter and \textbf{$M_{i}$} found for $L=0$ are
different from the ones found for $L=1$. 

\subsection{The three-body problem}

The baryon being a three-body system, so the problem is in the calculation
of the energy eigenvalue, which contains an integral of nine variables.
Before choosing the form of the test wave function, one defines the
coordinates of Jacobi $(\overrightarrow{\rho },\overrightarrow{\lambda })$,
which make it possible to pass from a three-body system to an equivalent
two-body system. These two relative coordinates represent respectively the
distance between the first two quarks ($q_{1,}q_{2}$), and the distance
between the third quark $q_{3}$ and the center of mass of the system $%
q_{1}q_{2}$. The Jacobi coordinates are defined as \cite{isgur4} :

\begin{eqnarray}
\ \overrightarrow{\rho } &=&\frac{1}{\sqrt{2}}(\overrightarrow{r_{1}}-%
\overrightarrow{r_{2}})  \label{rola} \\
\overrightarrow{\lambda } &=&\frac{1}{\sqrt{6}}(\overrightarrow{r_{1}}+%
\overrightarrow{r_{2}}-2\overrightarrow{r_{3}})  \notag
\end{eqnarray}

\bigskip Using these coordinates, we obtain the relative Hamiltonian $H_{R}$
written in the following form:

\begin{equation}
H_{R}=\frac{P_{\rho }^{2}}{2\mu _{\rho }}+\frac{P_{\lambda }^{2}}{2\mu
_{\lambda }}+M_{u}+\frac{m_{u}^{2}}{M_{u}}+\frac{M_{d}}{2}+\frac{m_{d}^{2}}{%
2M_{d}}+V(\overrightarrow{\rho },\overrightarrow{\lambda })  \label{(1.7)}
\end{equation}

With :

\hspace{2cm}$\mu _{\rho }=M_{u}$ \ \ \ \ \ \ \ \ \ ($u$ and $d$ represent
the quark flavors)

\hspace{2cm}$\mu _{\lambda }=\frac{3M_{u}M_{d}}{2M_{u}+M_{d}}$ \newline

The potential $V(\overrightarrow{\rho },\overrightarrow{\lambda })$ is
written in the form

\begin{eqnarray}
V(\overrightarrow{\rho },\overrightarrow{\lambda }) &=&-\frac{2\alpha _{s}}{3%
}\left( \frac{1}{\sqrt{2}\rho }+\frac{1}{\frac{1}{\sqrt{2}}\left\vert \sqrt{3%
}\overrightarrow{\lambda }+\overrightarrow{\rho }\right\vert }+\frac{1}{%
\frac{1}{\sqrt{2}}\left\vert \sqrt{3}\overrightarrow{\lambda }-%
\overrightarrow{\rho }\right\vert }\right)  \label{Vrola} \\
&&+\frac{\sigma }{2}\left( \sqrt{2}\rho +\frac{1}{\sqrt{2}}\left\vert \sqrt{3%
}\overrightarrow{\lambda }+\overrightarrow{\rho }\right\vert +\frac{1}{\sqrt{%
2}}\left\vert \sqrt{3}\overrightarrow{\lambda }-\overrightarrow{\rho }%
\right\vert \right) -2c  \notag
\end{eqnarray}

\section{Application of the model to the S-wave baryon}

In this part, one concentrates on the determination of the ground-states
masses of non-strange baryons with orbital momentum L=0. The parameter of
the wave function and masses ($M_u$, $M_d$) are determined using the
variational treatment.

\subsection{Determination of the proton wave function}

The total wave function $\left\vert qqq\right\rangle $ of the system is in
general constructed from the sum of $C_{A}\sum \chi \Psi \Phi $, where $%
\left( C_{A},\chi ,\Psi ,\Phi \right) $ represent respectively the color,
spin, spatial and flavor wave functions, with \textbf{$C_A$} totally
antisymmetric.

\begin{equation*}
\left\vert qqq\right\rangle \ =\left\vert Color\right\rangle _{A}\ \times
\left\vert Spatial,Spin,Flavor\right\rangle _{S}\ 
\end{equation*}

The indices $A$ and $S$ mean the antisymmetry and symmetry under the
exchange of any pair of quarks of equal masses.

\begin{eqnarray}
\left\vert N^{2}S_S\frac{1}{2}^+\right\rangle &=& C_A\Psi^S_{00} \frac{1}{%
\sqrt{2}}(\Phi^\rho_N\chi^\rho_{\frac{1}{2}}+\Phi^\lambda_N\chi^\lambda_{%
\frac{1}{2}})  \label{TWF00} \\
\left\vert \Delta^{4}S_S\frac{3}{2}^+\right\rangle &=& C_A\Phi^S_\Delta
\Psi^S_{00}\chi^S_{\frac{3}{2}}  \notag
\end{eqnarray}

One restricts oneself here to the spatial wave function for the calculation
carried out with the Hamiltonian $H_{R}$. To calculate the hyperfine
correction, one will take the total wave function defined in section $4$.
These corrections will give us the mass splitting between nucleon and $%
\Delta(1232) $. The spatial wave function $\Psi _{00}$ selected to describe
our system in the fundamental state $\left( \Psi _{LM}\longrightarrow L=0%
\text{, }M=0\right) $, is a Gaussian that one develops on the possible
states of the angular momenta of the systems $q_{1}q_{2}$ and $q_{3}$. These
states are indexed by $\mid l_{\rho }$, $l_{\lambda }\rangle$ with $%
\overrightarrow{L}=\overrightarrow{l_{\rho }}+\overrightarrow{l_{\lambda }}$
. The spatial wave function used in our calculation is in the following form:

\begin{eqnarray}
\Psi _{LM}(\overrightarrow{\rho },\overrightarrow{\lambda }) &=&\underset{%
l_{\rho },l_{\lambda }}{\sum }\underset{m_{\rho },m_{\lambda }}{\sum }%
C_{l_{\rho }l_{\lambda }}.N_{l_{\rho }l_{\lambda }}\left\langle l_{\rho
}m_{\rho }l_{\lambda }m_{\lambda }\mid LM\right\rangle \times \rho ^{l_{\rho
}}\lambda ^{l_{\lambda }}\   \label{PsiLM} \\
&&\times Exp[-\frac{1}{2}(\alpha _{\rho }^2\rho ^{2}+\alpha _{\lambda
}^2\lambda ^{2})]Y_{l_{\rho }}^{m_{\rho }}(\Omega _{\rho })Y_{l_{\lambda
}}^{m_{\lambda }}(\Omega _{\lambda })  \notag
\end{eqnarray}

where $C_{l_{\rho }l_{\lambda }}$ are coefficients determined by
diagonalisation of the $H_{R}$ matrix. The minimization of the system energy
allows the determination of the variational parameters of the spatial wave
function $(\alpha _{\rho },\alpha _{\lambda })$ and the dynamical masses $%
(M_{u},M_{d})$. The average value of the energy over the function $\Psi
_{00} $ is given by the following relation:

\begin{equation}
E(\alpha _{\rho },\alpha _{\lambda },M_{i})=\frac{\left\langle \Psi
_{00}\right\vert H_{R}\left\vert \Psi _{00}\right\rangle }{\left\langle \Psi
_{00}\mid \Psi _{00}\right\rangle }  \label{(1.10)}
\end{equation}

The only Clebsch-Gordan coefficients different from zero are those which
couple the orbital momentum associated to the Jacobian coordinates ($%
\overrightarrow{\rho },\overrightarrow{\lambda }$) with the total orbital
momentum $L=0$, and we have: 
\begin{equation}
|l_{\rho }-l_{\lambda }|\leq L=0\leq l_{\rho }+l_{\lambda }  \label{L}
\end{equation}%
The expression (\ref{L}) yields the constraint ($l_{\rho }=l_{\lambda }$).
So the states of this subsystem ($\overrightarrow{\rho },\overrightarrow{%
\lambda }$) which contribute to the spatial wave function have all the same
orbital momentum, starting with $l_{\rho }=l_{\lambda }=0$. From parity we
have $P=(-)^{L}=(-)^{l_{\rho }+l_{\lambda }}=+1$, so ($l_{\rho }+l_{\lambda
} $) is even. We do our treatment up to order $2$, i.e. $l_{\rho
}=l_{\lambda }=2$, since for higher orders the analytical calculation
becomes very complicated, moreover the contribution of the orbital momentum
terms $l>2$ is less important \cite{zouzou}. The total wave function must
obey the Pauli Exclusion Principle. In the baryon we have two identical
quarks, so the spatial wave function must be symmetrical in the exchange of
these two quarks $r_{1}\longleftrightarrow r_{2}$, this implies that in the
relative system, the function should be even in $\rho $ ($\overrightarrow{%
\rho }=\frac{1}{\sqrt{2}}(\overrightarrow{r_{1}}-\overrightarrow{r_{2}})$).
So in expression (\ref{PsiLM}), $l_{\rho }$ must be even.

Finally, by eliminating the contribution of the odd states in $l_{\rho }$,
only two terms contribute to the construction of the spatial wave function,
which is a superposition of $\left\vert l_{\rho }=0,l_{\lambda
}=0\right\rangle $ and $\left\vert l_{\rho }=2,l_{\lambda }=2\right\rangle $%
. So the calculation of energy reduces to the calculation of the matrix
elements of $H_{R}$ on the states $\left\vert l_\rho ,l_\lambda\right\rangle
= \left\vert 00\right\rangle$ and $\left\vert 22\right\rangle $.

If one notes the spatial wave function in the representation of Dirac, one
will have:

\begin{equation}
\left\vert \Psi _{00}\right\rangle =c_{1}\left\vert 00\right\rangle
+c_{2}\left\vert 22\right\rangle  \label{Psi00}
\end{equation}

Equation (\ref{Psi00}) shows that the physical state is a mixing of the
relative orbital momentum states ($l_{\rho }$,$l_{\lambda }=0,0$) and ($%
l_{\rho }$,$l_{\lambda }=2,2$), with mixing coefficients (c1, c2). To
calculate the energy of the physical state $\left\vert \Psi
_{00}\right\rangle $, one starts by evaluating the matrix elements of $H_{R}$
on the base \{$\left\vert 00\right\rangle $, $\left\vert 22\right\rangle $%
\}. The $2\times 2$ matrix is not diagonal. If one notes $T$ the kinetic
energy and $V$ the potential energy, the $H_R$ matrix is written:

\begin{equation}
H_{R}=\left( 
\begin{array}{cc}
H_R^{(00,00)} & H_R^{(00,22)} \\ 
H_R^{(22,00)} & H_R^{(22,22)}%
\end{array}
\right)=\left( 
\begin{array}{cc}
\left\langle 00\right\vert (T+V)\left\vert 00\right\rangle & \left\langle
00\right\vert V\left\vert 22\right\rangle \\ 
\left\langle 22\right\vert V\left\vert 00\right\rangle & \left\langle
22\right\vert (T+V)\left\vert 22\right\rangle%
\end{array}%
\right)  \label{HRm}
\end{equation}

It should be noted that by analogy with the work of Capstick \cite{capstick}%
, one has $\alpha _{\rho }$ $\simeq \alpha _{\lambda }$, because in the case
of the harmonic oscillator $\alpha _{\rho }$ and $\alpha _{\lambda }$ are
proportional to the masses $m_{u}$ and $m_{d}$ in the case of baryons with
the same quark masses. So one replaces in the function $\left\vert \Psi
_{00}\right\rangle $, $\alpha _{\rho }$ $=\alpha _{\lambda }=\alpha $. The
calculation of the elements $T_{l_{\rho }l_{\lambda },l_{\rho }^{\prime
}l_{\lambda }^{\prime }}$ is treated in detail in appendix (\ref{Cake}). The
non-zero elements are the two diagonal elements $T_{00,00}$ and $T_{22,22}$.
These elements are thus given according to ($\alpha $, $M_{u}$,$M_{d}$).

\begin{figure}[htb]
\centering%
\includegraphics[height=8cm]
{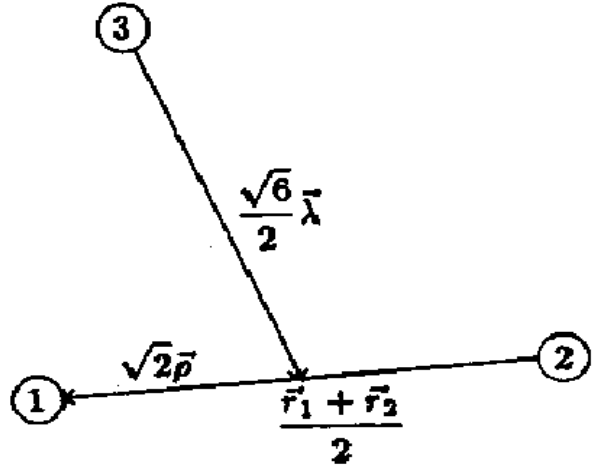}
\caption{Relative coordinates $\vec{\protect\rho}$ and $\vec{\protect\lambda}
$ }
\label{jacobi}
\end{figure}

The calculation of the matrix elements of the potential energy requires to
evaluate an integral of dimension six on the potential $V(\overrightarrow{%
\rho },\overrightarrow{\lambda })$ which contains terms coupled in ($\rho
,\lambda $) and the angle $\theta $ between the two vectors $(%
\overrightarrow{\rho },\overrightarrow{\lambda })$, see fig(\ref{jacobi}).


With :

\begin{eqnarray}
r_{12} &=&\sqrt{2}\rho  \label{(1.14)} \\
r_{13} &=&\frac{1}{\sqrt{2}}\sqrt{\rho ^{2}+\sqrt{3}\overrightarrow{\rho }.%
\overrightarrow{\lambda }+3\lambda ^{2}}  \notag \\
r_{23} &=&\frac{1}{\sqrt{2}}\sqrt{\rho ^{2}-\sqrt{3}\overrightarrow{\rho }.%
\overrightarrow{\lambda }+3\lambda ^{2}}  \notag
\end{eqnarray}

and using the form of \textquotedblleft Coulombian+ Linear" potential in
equation (\ref{Vrola})

one calculates now the matrix elements:

\begin{eqnarray}
\left\langle LM\right\vert V(\overrightarrow{\rho },\overrightarrow{\lambda }%
)\left\vert L^{\prime }M^{\prime }\right\rangle &=&\underset{m_{\rho
},m_{\lambda }}{\underset{l_{\rho },l_{\lambda }}{\dsum }}\underset{m_{\rho
}^{\prime },m_{\lambda }^{\prime }}{\underset{l_{\rho }^{\prime },l_{\lambda
}^{\prime }}{\dsum }}\left\langle l_{\rho }m_{\rho }l_{\lambda }m_{\lambda
}|LM\right\rangle ^{\ast }\left\langle l_{\rho }^{\prime }m_{\rho }^{\prime
}l_{\lambda }^{\prime }m_{\lambda }^{\prime }|L^{\prime }M^{\prime
}\right\rangle  \label{Vrl} \\
&&\times \dint d\overrightarrow{\rho }d\overrightarrow{\lambda }\rho
^{l_{\rho }+l_{\rho }^{\prime }}\lambda ^{l_{\lambda }+l_{\lambda }^{\prime
}}Exp\left[ -\alpha^2 (\rho ^{2}+\lambda ^{2})\right] V(\overrightarrow{\rho 
},\overrightarrow{\lambda })  \notag \\
&&\times Y_{l_{\rho }}^{m_{\rho }\ast }\left( \Omega _{\rho }\right)
Y_{l_{\lambda }}^{m_{\lambda }\ast }\left( \Omega _{\lambda }\right) \
Y_{l_{\rho }^{\prime }}^{m_{\rho }^{\prime }}\left( \Omega _{\rho }\right)
Y_{l_{\lambda }^{\prime }}^{m_{\lambda }^{\prime }}\left( \Omega _{\lambda
}\right)  \notag
\end{eqnarray}

To evaluate this integral in the coordinate system ($\overrightarrow{\rho },%
\overrightarrow{\lambda }$) , one

uses the hyperspherical coordinates ($\xi, \theta$) defined as follows :

\begin{eqnarray}
\rho &=&\xi \sin (\theta )  \label{(1.16)} \\
\lambda &=&\xi \cos (\theta )  \notag
\end{eqnarray}

This implies that : $\ \xi ^{2}=\rho ^{2}+\lambda ^{2}$ and $\theta =arctg(%
\frac{\rho }{\lambda })$, $0\leq \theta \leq \frac{\pi }{2}$

The integral (\ref{Vrl}) can be simplified in the following form:

\begin{equation}
V_{l_{\rho }l_{\lambda },l_{\rho }^{\prime }l_{\lambda }^{\prime
}}=\dint\limits_{0}^{\infty }d\xi \xi ^{5}\left( \frac{\xi }{2}\sigma
A_{l_{\rho }l_{\lambda },l_{\rho }^{\prime }l_{\lambda }^{\prime }}+\frac{2}{%
3\xi }\alpha _{s}B_{l_{\rho }l_{\lambda },l_{\rho }^{\prime }l_{\lambda
}^{\prime }}-2c\right) R_{l_{\rho }l_{\lambda }}(\xi )R_{l_{\rho }^{\prime
}l_{\lambda }^{\prime }}(\xi )  \label{Vrl2}
\end{equation}

With :

\begin{equation*}
\ R_{l_{\rho }l_{\lambda }}(\xi )=K_{l_{\rho }l_{\lambda }}\ \xi ^{l_{\rho
}+l_{\lambda }}Exp\left[ -\frac{1}{2}\alpha^2 \xi ^{2}\right]
\end{equation*}

$K_{l_{\rho }l_{\lambda }}$is a normalization factor. The problem now
amounts to the determination of elements $A_{l_{\rho }l_{\lambda },l_{\rho
}^{\prime }l_{\lambda }^{\prime }}$ and $B_{l_{\rho }l_{\lambda },l_{\rho
}^{\prime }l_{\lambda }^{\prime }}$. These elements represent the angular
part in the integral (\ref{Vrl}). The method carried out by calculation of $%
A_{l_{\rho }l_{\lambda },l_{\rho }^{\prime }l_{\lambda }^{\prime }}$ and $%
B_{l_{\rho }l_{\lambda },l_{\rho }^{\prime }l_{\lambda }^{\prime }}$ is
developed in detail in appendix (\ref{Cape}).

\subsection{Mass of the proton and mixing coefficients}

The analytic form of the Hamiltonian matrix elements of equation (\ref{HRm})
is: 
\begin{eqnarray}
H^{(00,00)}_R &=& \frac{1}{2}\{-1.74320+\frac{m_d^2}{M_d}+M_d+\frac{2m_u^2}{%
M_u}+2M_u  \notag \\
& &+\frac{0.737249}{\alpha}-2.74449\alpha+\alpha^2 (\frac{1}{M_d}+\frac{2}{%
M_u})\}  \label{H0000} \\
H^{(00,22)}_R &=& (0.04396-0.08154)\frac{1}{\alpha}  \label{H0022} \\
H^{(22,00)}_R &=& H^{(00,22)}_R  \label{2200} \\
H^{(22,22)}_R &=& -0.87160+\frac{0.5m_d^2}{M_d}+0.5M_d+\frac{m_u^2}{M_u}+M_u
\label{2222} \\
& &+\frac{0.114109}{\alpha}-0.181177\alpha+(\frac{1.16667}{M_d}+\frac{%
2.33333\alpha^2}{M_u})\alpha^2  \notag \\
\notag
\end{eqnarray}
After diagonalisation and minimization of the Hamiltonian matrix, we find
the following result (using an algorithm of Mathematica $5.0$ program) : 
\newline

$E_{min}^1=1068 MeV$ $\longleftarrow$ ($\alpha=200 MeV$, $M_u=M_d=480MeV$)

$E_{min}^2=1412 MeV$ $\longleftarrow$ ($\alpha=337 MeV$, $M_u=M_d=580MeV$) 
\newline
\newline
The physical state energy is the lowest value of these two eigenvalues ($%
E_{min}^1$,$E_{min}^2$). Table (\ref{swanpar}) gives the parameters of the
potential model considered in this work, which were used in \cite{swanson}.
The results of the computation of the nucleon and $\Delta(1232) $ mass are
summarized in table (\ref{promass}) as well as the values of the variational
parameters ($\alpha$, $M_u$, $M_d$). These parameters will be used in the
calculation of the corrections of the spin-spin type in the
semi-relativistic model (cf. section \ref{Hyper}).

\bigskip

It should be noted here that the mass of both nucleon and $\Delta(1232) $
appearing in table (\ref{promass}) is the energy of the physical state,
which is a mixing of the states $\left\vert 00\right\rangle $ and $%
\left\vert 22\right\rangle $, where the coefficients of the mixing obtained
after diagonalisation, are given by:

\begin{equation}
\left\vert N,\Delta \right\rangle =0.9761\left\vert 00\right\rangle
-0.2673\left\vert 22\right\rangle \longrightarrow E_{N,\Delta }=1068MeV
\label{(1.22)}
\end{equation}


\subsection{Hyperfine interaction and calculation of the baryons masses 
\label{Hyper}}

The potential of the hyperfine interaction is composed of two terms (\cite%
{isgur1},\cite{karl}) :

\begin{equation}
V_{hyp}=V_{c}+V_{t}  \label{Vhyp}
\end{equation}

Where the first term (called contact term of Fermi), represents the
spin-spin interaction between the quarks in the baryon. It is operative only
in the fundamental state, where the orbital momentum is equal to zero.

\begin{equation}
V_{c}=-\sum_{i<j=1}^{N}\alpha _{ij}\frac{8\pi \alpha _{h}}{3M_{i}M_{j}}\frac{%
\sigma _{h}^{3}}{\sqrt{\pi ^{3}}}\exp (-\sigma _{h}^{2}r_{ij}^{2})S_{i}.S_{j}
\label{Vcon}
\end{equation}

$(\alpha _{h},\sigma _{h})$ are parameters fitted in reference \cite{swanson}%
, (see table(1)). The second term (called tensor term) represents the static
interaction of two intrinsic magnetic dipoles. It is operative only if the
orbital momentum is larger than zero. It is written in the form:

\begin{equation}
V_{t}=\sum_{i<j=1}^{N}\alpha _{ij}\frac{\alpha _{s}}{M_{i}M_{j}}\frac{1}{%
r_{ij}^{3}}\left( 3(S_{i}.r)(S_{i}.r)-S_{i}.S_{j}\right)  \label{Vten}
\end{equation}

This term enables us to have the mixing coefficients of non-strange baryons (%
$S=0$) in the $P$ wave ($L=1$). The first term $V_{c}$ enables us to
separate between baryons of the same $J^{P}$ but of different spin. From
equations \{(\ref{TWF00}), (\ref{Vcon})\}, we obtain the mass of the nucleon
and $\Delta(1232)$ (mass splitting $\Delta -N$). Table (\ref{promass}) shows
the results obtained for the baryons masses, with and without hyperfine
correction.

\begin{center}
\begin{table}[tbp] \centering%
\begin{tabular}{ccccccc}
\hline
$\alpha _{h}$ & $\sigma _{h}$ & $\alpha _{s}$ & $\sigma $ & $\frac{3}{4}%
c(MeV)$ & $m_{u}(MeV)$ & $m_{d}(MeV)$ \\ \hline
0.840 & 0.700 & 0.857 & 0.154 & -436 & 375 & 375 \\ \hline\hline
\end{tabular}
\caption{Parameters of the quark potential model \cite{swanson}
\label{swanpar}}%
\end{table}%

\bigskip

\bigskip 
\begin{table}[tbp] \centering%
\begin{tabular}{|c|c|c|c|c|c|c|}
\hline
& $\alpha $ & $M_{u}$ & $M_{d}$ & $M_{0}$ & $M_{cor}$ & $Exp$ \\ \hline
$N(\frac{1}{2}^{+})$ & 200 & 480 & 480 & 1068 & 968 & 938 \\ 
$\Delta (\frac{3}{2}^{+})$ & 200 & 480 & 480 & 1068 & 1168 & 1232 \\ \hline
\end{tabular}%
\caption{The ground-state energy of baryons (in MeV). $M_0$ is the baryon
mass without hyperfine correction and $M_{cor}$ represents the baryon mass
with spin-spin ($V_c$) correction. \label{promass}}%
\end{table}%
\end{center}


\section{Application of the model to the P-wave baryon}

One carries out now the calculation of the first excited state energy of the
baryon with $L=1$ and negative parity, using the same method as for the
S-wave. In this case one will build the total wave function, which has to be
antisymmetric.

\subsection{Determination of the wave function for $L=1$}

In the construction of the spatial wave function, the only possible states
for $(l_{\rho },l_{\lambda })$ which allow to have an orbital momentum L=1
and negative parity are $(l_{\rho }=1,l_{\lambda }=0)$ and $(l_{\rho
}=0,l_{\lambda }=1)$ which are noted respectively ($\Psi _{1M}^{\rho }$, $%
\Psi _{1M}^{\lambda }$). Note from equation (\ref{rola}) that $\Psi
_{1M}^{\lambda }$ is even under the exchange of the first two quarks, while
the analogous wave function $\Psi _{1M}^{\rho }$ is odd, since $\rho $ and $%
\lambda $ respectively indicate mixed antisymmetry and mixed symmetry under
this transposition. These states form a representation of dimension two of
the permutation group $S_{3}$, which allows to exchange each pair of quarks
in the baryon.

These two functions have the following form:

\begin{eqnarray}
\Psi^\rho _{1M}(\overrightarrow{\rho },\overrightarrow{\lambda }) &=& C_{10}%
\underset{m_{\rho },m_{\lambda }}{\sum } N_{10}\left\langle 1m_{\rho
}0m_{\lambda }\mid 1M\right\rangle  \notag \\
& & \times \mathbf{\rho} Exp[-\frac{1}{2}\alpha^2 (\rho
^{2}+\lambda^{2})]Y_1^{m_{\rho }}(\Omega _{\rho })Y_0^{m_{\lambda }}(\Omega
_{\lambda })
\end{eqnarray}
\begin{eqnarray}
\Psi^\lambda _{1M}(\overrightarrow{\rho },\overrightarrow{\lambda }) &=&
C_{01}\underset{m_{\rho },m_{\lambda }}{\sum }N_{01}\left\langle 0m_{\rho
}1m_{\lambda }\mid 1M\right\rangle  \notag \\
& & \times \mathbf{\lambda} Exp[-\frac{1}{2}\alpha^2 (\rho
^{2}+\lambda^{2})]Y_0^{m_{\rho }}(\Omega _{\rho })Y_1^{m_{\lambda }}(\Omega
_{\lambda })  \notag
\end{eqnarray}

One will have also to consider the coupling of the spins $S$ with the
orbital momentum $L$ to build the total angular momentum $\overrightarrow{J}=%
\overrightarrow{L}+\overrightarrow{S}$. In short, here is the construction
of $J$:

The spin of the quark system : $S=\frac{1}{2},\frac{3}{2}$ , is constructed
as the triple product $\left( \frac{1}{2}\otimes \frac{1}{2}\otimes \frac{1}{%
2}\right). $

One has $L=1$, one thus distinguishes two constructions of $J$ for the
doublet and the quadruplet of spin :

$1)$ $S=\frac{1}{2}$, $L=1\longrightarrow $ $^{2}P_{\frac{1}{2}}$, $^{2}P_{%
\frac{3}{2}}$, \ \ \ \ $J^{P}=\frac{1}{2}^{-},\frac{3}{2}^{-}$

$2)$ $S=\frac{3}{2}$, $L=1\longrightarrow $ $^{4}P_{\frac{1}{2}}$, $^{4}P_{%
\frac{3}{2}}$, $^{4}P_{\frac{5}{2}}$,\ \ \ \ $J^{P}=\frac{1}{2}^{-},\frac{3}{%
2}^{-},\frac{5}{2}^{-}$

The nucleon states of total angular momentum $J=\frac{1}{2},\frac{3}{2}$ are
a mixing of the doublet and quadruplet of spin, for example:

\begin{eqnarray}
\left\vert J^{P}=(\frac{1}{2})^{-}\right\rangle &=&\alpha _{1}\left\vert
^{4}P_{\frac{1}{2}}\right\rangle +\beta _{1}\left\vert ^{2}P_{\frac{1}{2}%
}\right\rangle  \label{(1.26)} \\
\left\vert J^{P}=(\frac{3}{2})^{-}\right\rangle &=&\alpha _{2}\left\vert
^{4}P_{\frac{3}{2}}\right\rangle +\beta _{2}\left\vert ^{2}P_{\frac{3}{2}%
}\right\rangle  \notag
\end{eqnarray}

Thus, the wave function will be built as :

\begin{eqnarray}
\left\vert ^{2}P_{\frac{1}{2}}\right\rangle &\sim& \sum \Psi _{1M}\chi _{%
\frac{1}{2}}\Phi \\
\left\vert ^{4}P_{\frac{1}{2}}\right\rangle &\sim& \sum \Psi _{1M}\chi _{%
\frac{3}{2}}\Phi  \notag  \label{tofunc}
\end{eqnarray}
The construction of the spin-flavor wave function is carried out in analogy
with the notation of Karl and Isgur and collaborators (\cite{isgur2}, \cite%
{capstick}). The flavor-mixed antisymmetry and symmetry combinations of ``$%
uud$'' are:

\begin{eqnarray}
\Phi _{p}^{\rho } &=&\frac{1}{\sqrt{2}}(udu-duu)  \label{(1.27)} \\
\Phi _{p}^{\lambda } &=&-\frac{1}{\sqrt{6}}(duu+udu-2uud)  \notag
\end{eqnarray}

In the same way, the spin states are:

\begin{eqnarray}
\chi _{\frac{1}{2}}^{\rho } &=&\frac{1}{\sqrt{2}}(\left\vert \uparrow
\downarrow \uparrow \right\rangle -\left\vert \downarrow \uparrow \uparrow
\right\rangle )  \label{(1.28)} \\
\chi _{\frac{1}{2}}^{\lambda } &=&-\frac{1}{\sqrt{6}}(\left\vert \downarrow
\uparrow \uparrow \right\rangle +\left\vert \uparrow \downarrow \uparrow
\right\rangle -2\left\vert \uparrow \uparrow \downarrow \right\rangle ) 
\notag
\end{eqnarray}

One must add the spin wave function for the quadruplet which is completely
symmetric:

\begin{equation}
\chi _{\frac{3}{2}}^{S}=\left\vert \uparrow \uparrow \uparrow \right\rangle
\label{(1.29)}
\end{equation}

Thus, the total wave function will be built according to the following
combinations:

\begin{eqnarray}
\left\vert N^{2}P_M(\frac{1}{2}^-, \frac{3}{2}^-)\right\rangle &=& C_A\frac{1%
}{2}[\Phi^\rho_N(\Psi^\rho_{1M}\chi^\lambda_{\frac{1}{2}}+\Psi^\lambda_{1M}%
\chi^\rho_{\frac{1}{2}})+\Phi^\lambda_N(\Psi^\rho_{1M}\chi^\rho_{\frac{1}{2}%
}-\Psi^\lambda_{1M}\chi^\lambda_{\frac{1}{2}})]  \notag \\
\left\vert N^{4}P_M(\frac{1}{2}^-, \frac{3}{2}^-)\right\rangle &=&
C_A\chi^S_{\frac{3}{2}} \frac{1}{\sqrt{2}}(\Phi^\rho_N\Psi^\rho_{1M}+\Phi^%
\lambda_N\Psi^\lambda_{1M})  \label{TWF} \\
\left\vert \Delta^{2}P_M(\frac{1}{2}^-, \frac{3}{2}^-)\right\rangle &=&
C_A\Phi^S_\Delta\frac{1}{\sqrt{2}}(\Psi^\rho_{1M}\chi^\rho_{\frac{1}{2}%
}+\Psi^\lambda_{1M}\chi^\lambda_{\frac{1}{2}})  \notag
\end{eqnarray}

From equation (\ref{TWF}), one sees that we have two groups of states: $%
N^{\ast }$ and $\Delta ^{\ast }$. Each group has its own flavor wave
function.

\subsection{Excited nucleon mass}

For the P-wave excited-state nucleon, the Hamiltonian matrix is written as
follows:

\begin{equation}
H_{R}=\left( 
\begin{array}{cc}
\left\langle 01\right\vert H_R\left\vert 01\right\rangle & \left\langle
01\right\vert H_R\left\vert 10\right\rangle \\ 
\left\langle 10\right\vert H_R\left\vert 01\right\rangle & \left\langle
10\right\vert H_R\left\vert 10\right\rangle%
\end{array}%
\right)
\end{equation}

By the same method used in calculation of the baryons masses in the
ground-state $L=0$, one finds the elements of the $H_R$ matrix for $L=1$:

\begin{eqnarray}
\left\langle 01\right\vert H_R\left\vert 01\right\rangle &=& \frac{1}{2}%
(-1.7432+\frac{0.7372}{\alpha}-2.7445\alpha+\frac{m_d^2}{M_d}+M_d  \notag \\
& &+\alpha^2 (\frac{1}{M_d}+\frac{2}{M_u})+\frac{2m_u^2}{M_u})  \notag \\
\left\langle 01\right\vert H_R\left\vert 10\right\rangle &=& \left\langle
10\right\vert H_R\left\vert 01\right\rangle  \notag \\
&=& (0.0439-0.0815\alpha^2)\frac{1}{\alpha}  \notag \\
\left\langle 10\right\vert H_R\left\vert 10\right\rangle &=& -0.8716+\frac{%
0.1141}{\alpha}-\frac{0.1812}{\alpha}+\frac{1.1667\alpha^2}{M_d}+\frac{m_d^2%
}{2M_d}+\frac{M_d}{2}  \notag \\
& &+\frac{2.3333\alpha^2}{M_u}+\frac{m_u^2}{M_u}+M_u
\end{eqnarray}

After diagonalisation of this matrix, one obtains two eigenvalues of the
Hamiltonian $H_R$ :

\begin{eqnarray}
H_R^1 &=& \frac{1}{2}[-1.7432+\frac{m_d^2}{M_d}+M_d+\frac{2m_u^2}{M_u}+2M_u+%
\frac{0.2027}{\alpha}-1.0745\alpha  \notag \\
& &+(\frac{1.3333}{M_d}+\frac{2.6666}{M_u})\alpha^2-\{-0.2863+\frac{0.0310}{%
\alpha^2}-(\frac{0.0193}{M_d}+\frac{0.0193}{M_u})\alpha  \notag \\
& &+0.7850 \alpha^2-(\frac{0.1433}{M_d}+\frac{0.1433}{M_u})\alpha^{3}+(\frac{%
0.1111}{M_d^2}+\frac{0.1111}{M_u^2}+\frac{0.2222}{M_uM_d})\alpha^4\}^\frac{1%
}{2}]  \notag \\
H_R^2 &=& \frac{1}{2}[-1.7432+\frac{m_d^2}{M_d}+M_d+\frac{2m_u^2}{M_u}+2M_u+%
\frac{0.2027}{\alpha}-1.0745\alpha \\
& &+(\frac{1.3333}{M_d}+\frac{2.6666}{M_u})\alpha^2+\{-0.2863+\frac{0.0310}{%
\alpha^2}-(\frac{0.0193}{M_d}+\frac{0.0193}{M_u})\alpha  \notag \\
& &+0.7850 \alpha^2-(\frac{0.1433}{M_d}+\frac{0.1433}{M_u})\alpha^{3}+(\frac{%
0.1111}{M_d^2}+\frac{0.1111}{M_u^2}+\frac{0.2222}{M_uM_d})\alpha^4\}^\frac{1%
}{2}]  \notag
\end{eqnarray}

The minimization of these two eigenvalues yields the following result :

\begin{eqnarray}
H_R^1 \to 1590 \ MeV  \notag \\
H_R^2 \to 1755 \ MeV
\end{eqnarray}

The minimal value ($1594$ $MeV$) corresponds to the nucleon mass in the
first excited state $L=1$ for the values ($\alpha =192MeV,M_{u}=M_{d}=1485MeV
$). In order to separate between the nucleon states of different total
angular momentum, one will apply the hyperfine potential defined in equation
(\ref{Vhyp}) which is composed of two parts, the contact term and the tensor
term.

\subsection{Hyperfine correction}

The orbital excitation $L=1$ of the baryon is carried by only one pair of
quarks, while the two other pairs have zero orbital momentum. This
decomposition of orbital momentum enables us to note that the tensor part of
the hyperfine interaction is operational only in the quark-pair with L=1. On
the other hand, the two other pairs with orbital momentum equal to zero are
controlled by the contact term of the hyperfine interaction. One will
calculate first the contact part, then the tensor part.

\subsubsection{The elements of the contact matrix}

The following calculation was carried out for only one pair of quarks, then
the result was multiplied by three. Furthermore the result does not depend
on the total angular momentum. Recall that the total spin for our system
takes the value ($S=\frac{1}{2}$, $S=\frac{3}{2}$) and thus we are left to
calculate the elements of the following matrix :

\begin{equation}
\left( 
\begin{array}{cc}
\left\langle N^{4}P_M(J) \right\vert V_c \left\vert N^{4}P_M(J) \right\rangle
& \left\langle N^{4}P_M(J) \right\vert V_c \left\vert N^{2}P_M(J)
\right\rangle \\ 
\  & \  \\ 
\left\langle N^{2}P_M(J) \right\vert V_c \left\vert N^{4}P_M(J) \right\rangle
& \left\langle N^{2}P_M(J) \right\vert V_c \left\vert N^{2}P_M(J)
\right\rangle \\ 
& 
\end{array}
\right)
\end{equation}

Using equation (\ref{TWF}) one obtains:

\begin{eqnarray}
\left\langle N^{4}P_M(J) \right\vert V_c \left\vert N^{4}P_M(J)
\right\rangle & = & 3\left\langle N^{4}P_M(J) \right\vert V^{12}_c
\left\vert N^{4}P_M(J) \right\rangle  \notag \\
& = & \frac{3}{2}\{\left\langle \chi^s_{\frac{3}{2}} \Psi^\lambda_{1M}
\right\vert V^{12}_c \left\vert \chi^s_{\frac{3}{2}} \Psi^\lambda_{1M}
\right\rangle  \notag \\
& &\ \ \ +\left\langle \chi^s_{\frac{3}{2}} \Psi^\rho_{1M} \right\vert
V^{12}_c \left\vert \chi^s_{\frac{3}{2}} \Psi^\rho_{1M} \right\rangle \} 
\notag \\
& = &\frac{3}{2}\frac{8\pi \alpha _{h}}{3M^2}\frac{2}{3}\frac{\sigma _{h}^{3}%
}{\sqrt{\pi ^{3}}}\left\langle \chi^s_{\frac{3}{2}} \right\vert S_1.S_2
\left\vert \chi^s_{\frac{3}{2}} \right\rangle  \label{Vc44} \\
& &\times \{\left\langle \Psi^\lambda_{1M} \right\vert Exp (-\sigma
_{h}^{2}r_{12}^{2}) \left\vert \Psi^\lambda_{1M} \right\rangle  \notag \\
& &\ \ \ +\left\langle \Psi^\rho_{1M} \right\vert Exp (-\sigma
_{h}^{2}r_{12}^{2}) \left\vert \Psi^\rho_{1M} \right\rangle \}  \notag
\end{eqnarray}
\begin{eqnarray}
\left\langle N^{2}P_M(J) \right\vert V_c \left\vert N^{2}P_M(J)
\right\rangle & = & 3\left\langle N^{2}P_M(J) \right\vert V^{12}_c
\left\vert N^{2}P_M(J) \right\rangle  \notag \\
& = & \frac{3}{2}\frac{8\pi \alpha _{h}}{3M^2}\frac{2}{3}\frac{\sigma
_{h}^{3}}{\sqrt{\pi ^{3}}}  \notag \\
& &\left(\left\langle \chi^{\rho}_{\frac{1}{2}} \right\vert S_1.S_2
\left\vert \chi^{\rho}_{\frac{1}{2}} \right\rangle + \left\langle
\chi^{\lambda}_{\frac{1}{2}} \right\vert S_1.S_2 \left\vert \chi^{\lambda}_{%
\frac{1}{2}} \right\rangle \right)  \notag \\
& &\times (\left\langle \Psi^\lambda_{1M} \right\vert Exp (-\sigma
_{h}^{2}r_{12}^{2}) \left\vert \Psi^\lambda_{1M} \right\rangle  \notag \\
& &\ \ \ +\left\langle \Psi^\rho_{1M} \right\vert Exp (-\sigma
_{h}^{2}r_{12}^{2}) \left\vert \Psi^\rho_{1M} \right\rangle)  \label{Vc22}
\end{eqnarray}
After integration of the spatial part, and using the Jacobi variable change $%
(\mathbf{r_{12}=\sqrt{2}\rho})$ one obtains the following result :

\begin{eqnarray}
\left\langle \Psi^\lambda_{1M} \right\vert Exp (-\sigma _{h}^{2}r_{12}^{2})
\left\vert \Psi^\lambda_{1M} \right\rangle & = & \left\langle
\Psi^\lambda_{1M} \right\vert Exp (-2\sigma _{h}^{2}\rho^{2}) \left\vert
\Psi^\lambda_{1M} \right\rangle  \notag \\
& = & \frac{\alpha^3}{(\alpha^2 +2\sigma^2_h)^\frac{3}{2}}
\end{eqnarray}
\begin{eqnarray}
\left\langle \Psi^\rho_{1M} \right\vert Exp (-\sigma _{h}^{2}r_{12}^{2})
\left\vert \Psi^\rho_{1M} \right\rangle & = & \left\langle \Psi^\rho_{1M}
\right\vert Exp (-2\sigma _{h}^{2}\rho^{2}) \left\vert \Psi^\rho_{1M}
\right\rangle  \notag \\
& = & \frac{\alpha^5}{(\alpha^2 +2\sigma^2_h)^\frac{5}{2}}
\end{eqnarray}

Equations \{(\ref{Vc44}), (\ref{Vc22})\} are written then in the following
form:

\begin{eqnarray}
\left\langle N^{4}P_M(J) \right\vert V^{12}_c \left\vert N^{4}P_M(J)
\right\rangle &=&\frac{3\sigma^3_h}{(\alpha^2 +2\sigma^2_h)^\frac{3}{2}}%
\left(\frac{\alpha^2}{\alpha^2 +2\sigma^2_h}+1\right)D\frac{\alpha^3}{\sqrt{%
\pi}}  \notag \\
& &  \label{Vc12} \\
\left\langle N^{2}P_M(J) \right\vert V^{12}_c \left\vert N^{2}P_M(J)
\right\rangle &=&-\frac{3\sigma^3_h}{(\alpha^2 +2\sigma^2_h)^\frac{3}{2}}%
\left(\frac{\alpha^2}{\alpha^2 +2\sigma^2_h}+1\right)D\frac{\alpha^3}{\sqrt{%
\pi}}  \notag \\
\left\langle N^{4}P_M(J) \right\vert V^{12}_c \left\vert N^{2}P_M(J)
\right\rangle &=& \left\langle N^{2}P_M(J) \right\vert V^{12}_c \left\vert
N^{4}P_M(J) \right\rangle = 0  \notag
\end{eqnarray}
where $D=\frac{2\alpha_h}{3M^2}$. The result of equation (\ref{Vc12}) does
not depend on the total angular momentum J. The non-diagonal elements of the 
$V^{12}_c$ matrix are equal to zero. For the case of the $\Delta^*$, the
tensor part of the hyperfine potential does not contribute to the correction
of the mass, however the contact term gives:

\begin{eqnarray}
\left\langle \Delta^{2}P_M(J) \right\vert V_c \left\vert
\Delta^{2}P_M(J)\right\rangle&=& \left\langle \Delta^{2}P_M(J) \right\vert
V^{12}_c \left\vert \Delta^{2}P_M(J)\right\rangle  \notag \\
&=&\frac{3}{2}\{ \left\langle \Psi^{\rho}_{1M}\chi^{\rho}_\frac{1}{2}\
\right\vert V^{12}_c \left\vert \Psi^{\rho}_{1M}\chi^{\rho}_\frac{1}{2}
\right\rangle  \notag \\
& &\ \ \ +\left\langle \Psi^{\lambda}_{1M}\chi^{\lambda}_\frac{1}{2}\
\right\vert V^{12}_c \left\vert \Psi^{\lambda}_{1M}\chi^{\lambda}_\frac{1}{2}
\right\rangle \}  \notag \\
&=&\frac{3}{2}\frac{2\alpha_h}{3M^2}\frac{8\sigma^3_h}{3\sqrt{\pi}}  \notag
\\
& &\times [\left\langle \chi^{\rho}_\frac{1}{2}\ \right\vert \mathbf{S_1.S_2}
\left\vert \chi^{\rho}_\frac{1}{2} \right\rangle \left\langle
\Psi^{\rho}_{1M}\ \right\vert Exp(2\sigma^2_h\rho^2) \left\vert
\Psi^{\rho}_{1M} \right\rangle  \notag \\
& &\ \ \ +\left\langle \chi^{\lambda}_\frac{1}{2}\ \right\vert \mathbf{%
S_1.S_2} \left\vert \chi^{\lambda}_\frac{1}{2} \right\rangle \left\langle
\Psi^{\lambda}_{1M}\ \right\vert Exp(2\sigma^2_h\rho^2) \left\vert
\Psi^{\lambda}_{1M} \right\rangle]  \notag \\
&=&\frac{3}{2}\frac{8D\sigma^3_h}{3\sqrt{\pi}}\left(-\frac{3}{4}\frac{%
\alpha^3}{(\alpha^2 +3\sigma^2_h)^\frac{3}{2}}+\frac{1}{4}\frac{\alpha^5}{%
(\alpha^2 +3\sigma^2_h)^\frac{5}{2}} \right)  \notag \\
&=&\frac{3}{2}\frac{8D\sigma^3_h}{3\sqrt{\pi}}\frac{1}{4}\frac{\alpha^3}{%
(\alpha^2 +3\sigma^2_h)^\frac{3}{2}}\left(-3+\frac{\alpha^2}{(\alpha^2
+3\sigma^2_h)}\right)
\end{eqnarray}

\subsubsection{The elements of the tensor matrix}

The interaction between magnetic dipoles appears in the case $L=1$. The
elements of the tensor matrix depend on the total angular momentum $J$.
Using the formulas of reference \cite{isgur1}, we find the following results
:

\begin{eqnarray}
\left\langle N^{4}P_\frac{3}{2} \right\vert V^{12}_t \left\vert N^{4}P_\frac{%
3}{2} \right\rangle &=& -\frac{3D}{2}\left\langle \Psi^{\rho}_{11}
\right\vert \rho^{-3}(3cos^2\theta_\rho-1) \left\vert \Psi^{\rho}_{11}
\right\rangle  \notag \\
\left\langle N^{4}P_\frac{3}{2} \right\vert V^{12}_t \left\vert N^{2}P_\frac{%
3}{2} \right\rangle &=& \left\langle N^{2}P_\frac{3}{2} \right\vert V^{12}_t
\left\vert N^{4}P_\frac{3}{2} \right\rangle  \notag \\
&=& -\frac{3}{4}\left(\frac{5}{2}\right)^{\frac{1}{2}}\frac{D}{2}%
\left\langle \Psi^{\rho}_{11} \right\vert \rho^{-3}(3cos^2\theta_\rho-1)
\left\vert \Psi^{\rho}_{11} \right\rangle  \notag \\
\left\langle N^{2}P_\frac{3}{2} \right\vert V^{12}_t \left\vert N^{2}P_\frac{%
3}{2} \right\rangle&=& 0
\end{eqnarray}

\begin{eqnarray}
\left\langle N^{4}P_\frac{1}{2} \right\vert V^{12}_t \left\vert N^{4}P_\frac{%
1}{2} \right\rangle &=& \frac{15}{4}\frac{D}{2}\left\langle \Psi^{\rho}_{11}
\right\vert \rho^{-3}(3cos^2\theta_\rho-1) \left\vert \Psi^{\rho}_{11}
\right\rangle  \notag \\
\left\langle N^{4}P_\frac{1}{2} \right\vert V^{12}_t \left\vert N^{2}P_\frac{%
1}{2} \right\rangle &=& \left\langle N^{2}P_\frac{1}{2} \right\vert V^{12}_t
\left\vert N^{4}P_\frac{1}{2} \right\rangle  \notag \\
&=& \frac{15}{4}\frac{D}{2}\left\langle \Psi^{\rho}_{11} \right\vert
\rho^{-3}(3cos^2\theta_\rho-1) \left\vert \Psi^{\rho}_{11} \right\rangle 
\notag \\
\left\langle N^{2}P_\frac{1}{2} \right\vert V^{12}_t \left\vert N^{2}P_\frac{%
1}{2} \right\rangle&=& 0  \notag
\end{eqnarray}

with : 
\begin{eqnarray}
\left\langle \Psi^{\rho}_{11} \right\vert \rho^{-3}(3cos^2\theta_\rho-1)
\left\vert \Psi^{\rho}_{11} \right\rangle &=& -\frac{8}{15}\frac{\alpha ^{%
\frac{3}{2}}}{\sqrt{\pi}}
\end{eqnarray}

\subsubsection{Energy spectrum and mixing angles of the first excited states
of the baryons}

After having calculated the matrix elements of the hyperfine correction, the
total matrix is then written as follows: \newline
\newline
For $J=\frac{1}{2}$: 
\begin{equation}
V_{hyp}=D\frac{\alpha^3}{\sqrt{\pi}}\left( 
\begin{array}{cc}
\frac{3\sigma_{h}^3}{(\alpha^2+2\sigma_{h}^2)^\frac{3}{2}}\left(\frac{%
\alpha^2}{\alpha^2+2\sigma_{h}^2}+1\right)+1 & 1 \\ 
1 & -\frac{3\sigma_{h}^3}{(\alpha^2+2\sigma_{h}^2)^\frac{3}{2}}\left(\frac{%
\alpha^2}{\alpha^2+2\sigma_{h}^2}+1\right)%
\end{array}
\right)  \label{Vh1}
\end{equation}

For $J=\frac{3}{2}$: 
\begin{equation}
V_{hyp}=D\frac{\alpha^3}{\sqrt{\pi}}\left( 
\begin{array}{cc}
\frac{3\sigma_{h}^3}{(\alpha^2+2\sigma_{h}^2)^\frac{3}{2}}\left(\frac{%
\alpha^2}{\alpha^2+2\sigma_{h}^2}+1\right)+\frac{4}{5} & \frac{1}{\sqrt{10}}
\\ 
\frac{1}{\sqrt{10}} & -\frac{3\sigma_{h}^3}{(\alpha^2+2\sigma_{h}^2)^\frac{3%
}{2}}\left(\frac{\alpha^2}{\alpha^2+2\sigma_{h}^2}+1\right)%
\end{array}
\right)  \label{Vh2}
\end{equation}

After diagonalisation of these two matrices, one finds the following result
(in MeV):\newline
\newline
For $J=\frac{1}{2}$: \newline
\newline
\begin{equation}
(H_R+H_hyp)\left(%
\begin{array}{c}
N_{\frac{3}{2}\frac{1}{2}}^\star \\ 
N_{\frac{1}{2}\frac{1}{2}}^\star%
\end{array}%
\right) =\left(%
\begin{array}{cc}
1590-26 & 0 \\ 
0 & 1590+10%
\end{array}%
\right) \left(%
\begin{array}{c}
N_{\frac{3}{2}\frac{1}{2}}^\star \\ 
N_{\frac{1}{2}\frac{1}{2}}^\star%
\end{array}%
\right)
\end{equation}
\newline
\newline
For $J=\frac{3}{2}$: \newline
\newline
\begin{equation}
(H_R+H_hyp)\left(%
\begin{array}{c}
N_{\frac{3}{2}\frac{3}{2}}^\star \\ 
N_{\frac{1}{2}\frac{3}{2}}^\star%
\end{array}%
\right) =\left(%
\begin{array}{cc}
1590-17 & 0 \\ 
0 & 1590+30%
\end{array}%
\right) \left(%
\begin{array}{c}
N_{\frac{3}{2}\frac{3}{2}}^\star \\ 
N_{\frac{1}{2}\frac{3}{2}}^\star%
\end{array}%
\right)
\end{equation}
\newline
\newline

For the $\Delta^*$ one obtains : 
\begin{equation}
(H_R+H_hyp)|\Delta>=(1590+17) |\Delta>
\end{equation}
\newline
A physical state of total angular momentum $J$ is defined as a mixing of
states of spin momenta ($S=\frac{1}{2}$ and $\frac{3}{2}$). The mixing
angles are defined in the following equation, see \cite{karl} : 
\begin{equation}
|N_J^*(min)> = -\sin\theta \ |^4P_J> +\cos\theta \ |^2P_J>  \notag
\end{equation}
\begin{equation}
|N_J^*(max)> = \ \ \cos\theta \ |^4P_J> +\sin\theta \ |^2P_J>
\end{equation}
\newline
This gives the following result :

For $J=\frac{1}{2}$ : 
\begin{equation}
|N_{\frac{1}{2}}^*(min)> = 0.512 \ |^4P_\frac{1}{2}> +0.859 \ |^2P_\frac{1}{2%
}>  \notag
\end{equation}
\begin{equation}
|N_{\frac{1}{2}}^*(max)> = \ \ 0.859 \ |^4P_\frac{1}{2}> -0.512 \ |^2P_\frac{%
1}{2}>
\end{equation}
\newline
From these equations we obtain the mixing angle noted $\theta_s$ : \newline
\begin{equation}
\tan\theta_s =-\frac{0.512}{0.859} \ \ \ \Longrightarrow \theta_s=-30.8^\circ
\label{ths}
\end{equation}

For $J=\frac{3}{2}$ : 
\begin{equation}
|N_{\frac{3}{2}}^*(min)> = -0.107 \ |^4P_\frac{3}{2}> +0.994 \ |^2P_\frac{3}{%
2}>  \notag
\end{equation}
\begin{equation}
|N_{\frac{3}{2}}^*(max)> = \ \ 0.994 \ |^4P_\frac{3}{2}> +0.107 \ |^2P_\frac{%
3}{2}>
\end{equation}
\newline
From these equations we obtain the mixing angle noted $\theta_d$ : \newline
\begin{equation}
\tan\theta_d =\frac{0.107}{0.994} \ \ \ \Longrightarrow \theta_d=6.1^\circ
\label{thd}
\end{equation}
\newline

Without hyperfine correction we have found the mass of the excited nucleon ($%
L=1$) equal to ($M_N^*=1590MeV$) for the following values ($\alpha=192MeV,
M_u=M_d=1485MeV$).

\begin{center}
\begin{table}[hbt]
\begin{tabular}{ccccc}
\hline\hline
& $M^{\prime}_{cor}$ & $M^{\prime}_{cor}$ & $a_X$ & $a_X$ \\ 
& ``Coulombian + linear'' & \cite{isgur2} & ``Coulombian + linear'' & \cite%
{isgur2} \\ \hline
$N^{\star}_\frac{1}{2}(1535)$ & 1564 & 1490 & 0.859 & 0.850 \\ \hline
$N^{\star}_\frac{1}{2}(1650)$ & 1600 & 1655 & -0.512 & -0.530 \\ \hline
$N^{\star}_\frac{3}{2}(1520)$ & 1573 & 1535 & 0.994 & 0.990 \\ \hline
$N^{\star}_\frac{3}{2}(1700)$ & 1620 & 1745 & 0.107 & 0.110 \\ \hline
$\Delta^{\star}_\frac{1}{2}(1620)$ & 1607 & 1685 & 1 & 1 \\ \hline
$\Delta^{\star}_\frac{3}{2}(1700)$ & 1607 & 1685 & 1 & 1 \\ \hline\hline
\end{tabular}%
\caption{Masses $M^{\prime}_{cor}$(in $MeV$) and mixing angles $a_x$ of
baryons for L=1 with hyperfine correction calculated by ``Coulombian +
linear'' potential model and compared with Isgur and Karl result 
\protect\cite{isgur2}. }
\label{key}
\end{table}
\end{center}

Table (3) contains the nucleon masses ($M^{\prime}_{cor}$) and the mixing
angles ($a_X$) with hyperfine correction for ``Coulombian + linear''
potential model, compared with the work of Isgur and Karl \cite{isgur2} for $%
L=1$.\newline
The hyperfine correction to the energy spectrum of the excited nucleon is
small compared to that found in \cite{isgur2}.\newline
We remark that for the nucleon with total angular momentum $J=\frac{1}{2}$,
one finds from equation (\ref{ths}) a mixing angle $\theta_s=-30.8^\circ$,
which is very close to the result of \cite{isgur2} $\theta_s=-31.7^\circ$.
For the nucleon with total angular momentum $J=\frac{3}{2}$, one finds from
equation (\ref{thd}) a mixing angle $\theta_d=6.1^\circ$, which is very
close to the result of \cite{isgur2} $\theta_d=6.3^\circ$. \newline
Equations (\ref{Vh1}, \ref{Vh2}) allow us to have the mixing angles $a_X$.
They depend on the spatial wave function parameter $\alpha$ and the
potential parameter $\sigma_h$ contrary to the work of Isgur and Karl \cite%
{isgur1}, where the $a_X$ are independent of any choice of parameters.

\section{Conclusion}

The choice of the potential used (Coulombian+linear) in our model and the
addition of the semi-relativistic correction of the kinetic part of the
Hamiltonian allowed us to obtain a good result for the masses of the
non-strange baryons in S-wave. We separated the nucleon state from the $%
\Delta (1232)$ using the contact potential ($M_{\Delta }-M_{N}=200MeV$).
This difference in mass between the nucleon and the $\Delta (1232)$ is small
compared to that found by Isgur and Karl \cite{isgur5} ($M_{\Delta
}-M_{N}=260MeV$). \newline
For $L=1$ we obtained the excited nucleon mass without hyperfine correction (%
$M_{N^{\ast }}=1590MeV$). The result is very close to that of reference \cite%
{isgur2}. On the other hand the hyperfine correction enables us to separate
between the states of the excited nucleon. This correction is very small
compared to the result of \cite{isgur2}. The difference is due to the
dynamical mass used in the hyperfine correction which is heavier than the
quark constituent masses ($M_{u}=M_{d}=480MeV\gg m_{u}=m_{d}=375MeV$ for $L=0
$) and ($M_{u}=M_{d}=1485MeV\gg m_{u}=m_{d}=375MeV$ for $L=1$).\newline
For the case of the mixing angles we obtained the same result as found by
Isgur and Karl \cite{isgur2}, on the other hand our elements of the
hyperfine potential matrix depend on the parameter $\alpha $ of the spatial
wave function and the parameter $\sigma $ of the linear part of the
potential.\newline

{{\Large \textbf{Acknowledgements}}}

We are grateful to H. Fonvieille for proofreading of the manuscript. We
would like to thank G. Karl, E. Swanson, S. Capstick and J.M. Richard for
fruitful discussions.

\newpage

\section{Appendix}

\subsection{Wave function form and development}

One takes a function of the ``Gaussian development'' :

\begin{eqnarray}
\Psi _{LM}(\overrightarrow{\rho },\overrightarrow{\lambda }) &=&\underset{%
l_{\rho },l_{\lambda }}{\sum }\underset{m_{\rho },m_{\lambda }}{\sum }%
C_{l_{\rho }l_{\lambda }}.N_{l_{\rho }l_{\lambda }}\left\langle l_{\rho
}m_{\rho }l_{\lambda }m_{\lambda }\mid LM\right\rangle \times \rho ^{l_{\rho
}}\lambda ^{l_{\lambda }}\   \label{(2.1)} \\
&&\times Exp[-\frac{1}{2}(\alpha _{\rho }^2\rho ^{2}+\alpha _{\lambda
}^2\lambda ^{2})]Y_{l_{\rho }}^{m_{\rho }}(\Omega _{\rho })Y_{l_{\lambda
}}^{m_{\lambda }}(\Omega _{\lambda })  \notag
\end{eqnarray}

\bigskip

$\left\langle l_{\rho }m_{\rho }l_{\lambda }m_{\lambda }\mid LM\right\rangle 
$ are Clebsch-Gordan coefficients (one couples the orbital momenta
associated to the Jacobian variables $(\overrightarrow{\rho },%
\overrightarrow{\lambda })$ with a total momentum $\overrightarrow{L}=%
\overrightarrow{l_{\rho }}+\overrightarrow{l_{\lambda }}$ ).

In the case of baryons with at least two identical quarks, the above
expression of $|\Psi _{LM}\rangle $ must satisfy the constraints imposed by
the Pauli principle. If the two identical quarks are numbered 1 and 2, their
distance is expressed by the Jacobi variable $\overrightarrow{\rho }$ , and
the spatial wave function must be even in $\rho $. Moreover the parity of
the proton is positive, and according to the following relation $%
(P=(-1)^{L}) $, one deduces that the total orbital momentum must be even.
Parameters of the spatial wave function $(\alpha _{\rho },\alpha _{\lambda })
$ are given by minimizing the total energy $E$ of our system $\left( \frac{%
\partial E}{\partial \alpha }\right) $. For that, one will calculate the
kinetic energy and the average potential energy of the system.

\ 

\subsection{Calculation of the average kinetic energy}

\label{Cake}

The average value of our kinetic energy is written as follows:

\ \ 

\begin{eqnarray}
\left\langle \Psi _{LM}\left\vert T\right\vert \Psi _{L^{\prime }M^{\prime
}}\right\rangle &=&\underset{m_{\rho },m_{\lambda }}{\underset{l_{\rho
},l_{\lambda }}{\dsum }}\underset{m_{\rho }^{\prime },m_{\lambda }^{\prime }}%
{\underset{l_{\rho }^{\prime },l_{\lambda }^{\prime }}{\dsum }}N_{l_{\rho
}l_{\lambda }}N_{l_{\rho }^{\prime }l_{\lambda }^{\prime }}\left\langle
l_{\rho }m_{\rho }l_{\lambda }m_{\lambda }|LM\right\rangle ^{\ast }
\label{Tc} \\
&&\times \left\langle l_{\rho }^{\prime }m_{\rho }^{\prime }l_{\lambda
}^{\prime }m_{\lambda }^{\prime }|L^{\prime }M^{\prime }\right\rangle \times
\dint d\overrightarrow{\rho }d\overrightarrow{\lambda }\rho ^{l_{\rho
}}\lambda ^{l_{\lambda }}  \notag \\
&&\times Exp\left[ -\frac{1}{2}(\alpha _{\rho }^2\rho^{2}+\alpha _{\lambda
}^2\lambda ^{2})\right]Y_{l_{\rho }}^{m_{\rho }\ast }\left( \Omega _{\rho
}\right) Y_{l_{\lambda }}^{m_{\lambda }\ast }\left( \Omega _{\lambda }\right)
\notag \\
&&\times \left[ \frac{P_{\rho }^{2}}{2\mu _{\rho }}+\frac{P_{\lambda }^{2}}{%
2\mu _{\lambda }} +M^{\prime }\right] \ \rho ^{l_{\rho }^{\prime }}\lambda
^{l_{\lambda}^{\prime }}  \notag \\
&&\times \ Exp\left[ -\frac{1}{2}(\alpha _{\rho }^2\rho ^{2}+\alpha
_{\lambda }^2\lambda ^{2})\right] Y_{l_{\rho }^{\prime }}^{m_{\rho }^{\prime
}}\left( \Omega _{\rho }\right) Y_{l_{\lambda }^{\prime }}^{m_{\lambda
}^{\prime }}\left( \Omega _{\lambda }\right)  \notag
\end{eqnarray}

With

$N_{l_{\rho }l_{\lambda }}=2\frac{\alpha _{\rho \lambda }^{l_{\rho
}+l_{\lambda }+3}}{\sqrt{\Gamma (\frac{3}{2}+l_{\rho })\Gamma (\frac{3}{2}%
+l_{\lambda })}}$

\bigskip $P_{r}^{2}=-\Delta _{r}=-\left[ \frac{1}{r^{2}}\frac{\partial }{%
\partial r}\left( r^{2}\frac{\partial }{\partial r}\right) -\frac{L_{r}^{2}}{%
r^{2}}\right] $ , \ \ \ \ \ \ \ $L_{r}^{2}$ $Y_{l_{r}}^{m_{r}}\left( \Omega
_{r}\right) =l_{r}(l_{r}+1)Y_{l_{r}}^{m_{r}}\left( \Omega _{r}\right) $

$M^{\prime }=M_{u}+\frac{m_{u}^{2}}{M_{u}}+\frac{M_{d}}{2}+\frac{m_{d}^{2}}{%
2M_{d}},$ \ \ \ \ \ \ $\mu _{\rho }=M_{u}$ \ , \ \ \ \ \ $\mu _{\lambda }=%
\frac{3M_{u}M_{d}}{2M_{u}+M_{d}}$

\bigskip

and :

\begin{eqnarray}
\frac{\partial }{\partial r}\left( r^{2}\frac{\partial }{\partial r}\left(
r^{l_{r}}Exp\left[ -\frac{1}{2}\alpha^2 r^{2}\right] \right) \right)
&=&\left( l_{r}(l_{r}+1)-\alpha^2 (2l_{r}+3)r^{2}+\alpha ^{4}r^{4}\right) 
\notag \\
&&\times r^{l_{r}}Exp\left[ -\frac{1}{2}\alpha^2 r^{2}\right]  \notag
\end{eqnarray}

We can now resolve equation (\ref{Tc}) as follows :

\begin{eqnarray}
\left\langle \Psi _{LM}\left\vert T\right\vert \Psi _{L^{\prime }M^{\prime
}}\right\rangle &=&\underset{m_{\rho },m_{\lambda }}{\underset{l_{\rho
},l_{\lambda }}{\dsum }}\underset{m_{\rho }^{\prime },m_{\lambda }^{\prime }}%
{\underset{l_{\rho }^{\prime },l_{\lambda }^{\prime }}{\dsum }N_{l_{\rho
}l_{\lambda }}N_{l_{\rho }^{\prime }l_{\lambda }^{\prime }}}\left\langle
l_{\rho }m_{\rho }l_{\lambda }m_{\lambda }|LM\right\rangle ^{\ast } \\
&&\times \left\langle l_{\rho }^{\prime }m_{\rho }^{\prime }l_{\lambda
}^{\prime }m_{\lambda }^{\prime }|L^{\prime }M^{\prime }\right\rangle \times 
\left[ R\left( P_{\rho }^{2}\right) +R\left( P_{\lambda }^{2}\right)
+M^{\prime \prime }-L^{2}\right]  \notag
\end{eqnarray}

With:

\bigskip

\begin{eqnarray}
R\left( P_{\rho }^{2}\right) &=&\frac{1}{2\mu _{\rho }}\int d\rho d\lambda
\left( \lambda ^{l_{\lambda }+l_{\lambda }^{\prime }+2}Exp\left[ -\alpha
_{\lambda }^2\lambda ^{2}\right] \right)  \label{(2.5)} \\
&&\times \left( l_{\rho }^{\prime }(l_{\rho }^{\prime }+1)-\alpha _{\rho
}^2(2l_{\rho }^{\prime }+3)\rho ^{2}+\alpha _{\rho }^{4}\rho ^{4}\right) 
\notag \\
&&\times \rho ^{l_{\rho }+l_{\rho }^{\prime }}Exp\left[ -\alpha _{\rho
}^2\rho ^{2}\right] \delta _{l_{\rho }l_{\rho }^{\prime }}\delta _{m_{\rho
}m_{\rho }^{\prime }}\delta _{l_{\lambda }l_{\lambda }^{\prime }}\delta
_{m_{\lambda }m_{\lambda }^{\prime }}  \notag
\end{eqnarray}

$\bigskip $

\begin{eqnarray}
R\left( P_{\lambda }^{2}\right) &=&\frac{1}{2\mu _{\lambda }}\int d\rho
d\lambda \left( \rho ^{l_{\rho }+l_{\rho }^{\prime }+2}Exp\left[ -\alpha
_{\rho }^2\rho ^{2}\right] \right)  \label{(2.6)} \\
&&\times \left( l_{\lambda }^{\prime }(l_{\lambda }^{\prime }+1)-\alpha
_{\lambda }^2(2l_{\lambda }^{\prime }+3)\lambda ^{2}+\alpha _{\lambda
}^{4}\lambda ^{4}\right)  \notag \\
&&\times \lambda ^{l_{\lambda }+l_{\lambda }^{\prime }}Exp\left[ -\alpha
_{\lambda }^2\lambda ^{2}\right] \delta _{l_{\rho }l_{\rho }^{\prime
}}\delta _{m_{\rho }m_{\rho }^{\prime }}\delta _{l_{\lambda }l_{\lambda
}^{\prime }}\delta _{m_{\lambda }m_{\lambda }^{\prime }}  \notag
\end{eqnarray}

$\bigskip $

\begin{eqnarray}
M^{\prime \prime } &=&\left( M_{u}+\frac{m_{u}^{2}}{M_{u}}+\frac{M_{d}}{2}+%
\frac{m_{d}^{2}}{2M_{d}}\right) \int d\rho d\lambda \left( \rho ^{l_{\rho
}+l_{\rho }^{\prime }+2}Exp\left[ -\alpha _{\rho }^2\rho ^{2}\right] \right)
\label{(2.7)} \\
&&\times \left( \lambda ^{l_{\lambda }+l_{\lambda }^{\prime }+2}Exp\left[
-\alpha _{\lambda }^2\lambda ^{2}\right] \right) \delta _{l_{\rho }l_{\rho
}^{\prime }}\delta _{m_{\rho }m_{\rho }^{\prime }}\delta _{l_{\lambda
}l_{\lambda }^{\prime }}\delta _{m_{\lambda }m_{\lambda }^{\prime }}  \notag
\end{eqnarray}

\bigskip

\begin{eqnarray}
L^{2} &=&\int d\rho d\lambda \left( \frac{l_{\rho }^{\prime }(l_{\rho
}^{\prime }+1)}{2\mu _{\rho }\rho ^{2}}+\frac{l_{\lambda }^{\prime
}(l_{\lambda }^{\prime }+1)}{2\mu _{\lambda }\lambda ^{2}}\right) \left(
\rho ^{l_{\rho }+l_{\rho }^{\prime }+2}Exp\left[ -\alpha _{\rho }^2\rho ^{2}%
\right] \right)  \label{(2.8)} \\
&&\times \left( \lambda ^{l_{\lambda }+l_{\lambda }^{\prime }+2}Exp\left[
-\alpha _{\lambda }^2\lambda ^{2}\right] \right) \delta _{l_{\rho }l_{\rho
}^{\prime }}\delta _{m_{\rho }m_{\rho }^{\prime }}\delta _{l_{\lambda
}l_{\lambda }^{\prime }}\delta _{m_{\lambda }m_{\lambda }^{\prime }}  \notag
\end{eqnarray}

\subsection{Calculation of the average potential energy}

\label{Cape}

As a first approximation, one takes $\left( \alpha _{\rho }=\alpha _{\lambda
}=\alpha \right) $ . The ``Coulombian + linear'' potential is written
according to the variables of Jacobi in equation (\ref{Vrola}).\newline
In order to calculate the average potential energy $V_{l_{\rho }l_{\lambda
},l_{\rho }^{\prime }l_{\lambda }^{\prime}}$, we are going to determine the
following coefficients $\left( A_{l_{\rho }l_{\lambda },l_{\rho }^{\prime
}l_{\lambda }^{\prime}},B_{l_{\rho }l_{\lambda },l_{\rho }^{\prime
}l_{\lambda }^{\prime}}\right) $ which represent the angular parts of this
integral. Using the hyperspherical coordinates $\left( \overrightarrow{\xi }%
,\theta\right)$, one finds the following result :

\begin{eqnarray}
\ A_{l_{\rho }l_{\lambda },l_{\rho }^{\prime }l_{\lambda }^{\prime }}&=&%
\frac{1}{\xi }\dint d\Omega _{5}\left( \sqrt{2}\rho +\frac{1}{\sqrt{2}}%
\left\vert \sqrt{3}\overrightarrow{\lambda }+\overrightarrow{\rho }%
\right\vert +\frac{1}{\sqrt{2}}\left\vert \sqrt{3}\overrightarrow{\lambda }-%
\overrightarrow{\rho }\right\vert \right)  \notag \\
& & \ \ \times Y_{l_{\rho }l_{\lambda }}^{m_{\rho }m_{\lambda }\ast }\left(
\Omega _{5}\right) Y_{l_{\rho }^{\prime }l_{\lambda }^{\prime }}^{m_{\rho
}^{\prime }m_{\lambda }^{\prime }}\left( \Omega _{5}\right)  \label{(1.18)}
\end{eqnarray}

\begin{eqnarray}
B_{l_{\rho }l_{\lambda },l_{\rho }^{\prime }l_{\lambda }^{\prime }}&=&-\xi
\dint d\Omega _{5}\left( \frac{1}{\sqrt{2}\rho }+\frac{1}{\frac{1}{\sqrt{2}}%
\left\vert \sqrt{3}\overrightarrow{\lambda }+\overrightarrow{\rho }%
\right\vert }+\frac{1}{\frac{1}{\sqrt{2}}\left\vert \sqrt{3}\overrightarrow{%
\lambda }-\overrightarrow{\rho }\right\vert }\right)  \notag \\
& & \ \ \ \times Y_{l_{\rho }l_{\lambda }}^{m_{\rho }m_{\lambda }\ast
}\left( \Omega _{5}\right) Y_{l_{\rho }^{\prime }l_{\lambda }^{\prime
}}^{m_{\rho }^{\prime }m_{\lambda }^{\prime }}\left( \Omega _{5}\right)
\label{(1.19)}
\end{eqnarray}

Where :

\ $\ \ \ \ \ \ \ d\Omega _{5}=d\Omega _{\rho }d\Omega _{\lambda }\sin
^{2}\theta \cos ^{2}\theta d\theta $ \ \ , \ \ \ \ $d\Omega _{i}\equiv \sin
\theta _{i}d\theta _{i}d\varphi _{i}$

\ \ \ \ \ \ \ $Y_{l_{\rho }l_{\lambda }}^{m_{\rho }m_{\lambda }}\left(
\Omega _{5}\right) =D_{l_{\rho }l_{\lambda }}\sin ^{l_{\rho }}\theta $\ $%
\cos ^{l_{\lambda }}\theta $\ $Y_{l_{\rho }}^{m_{\rho }}\left( \Omega _{\rho
}\right) Y_{l_{\lambda }}^{m_{\lambda }}\left( \Omega _{\lambda }\right) $ 
\newline

For that we use the following definition :

\bigskip 
\begin{equation}
B_{l_{\rho }l_{\lambda },l_{\rho }^{\prime }l_{\lambda }^{\prime }}=-\left[ 
\frac{1}{\sqrt{2}}\dint\limits_{0}^{\frac{\pi }{2}}d\theta \sin \theta \cos
^{2}\theta \varphi _{^{l_{\rho }l_{\lambda }}}^{\ast }\left( \theta \right)
\varphi _{^{l_{\rho }^{\prime }l_{\lambda }^{\prime }}}\left( \theta \right)
+\xi \left( I_{b}^{+}+I_{b}^{-}\right) \right]  \label{(2.15)}
\end{equation}

\begin{equation}
A_{l_{\rho }l_{\lambda },l_{\rho }^{\prime }l_{\lambda }^{\prime }}=\left[ 
\sqrt{2}\dint\limits_{0}^{\frac{\pi }{2}}d\theta \sin \theta \cos ^{2}\theta
\varphi _{^{l_{\rho }l_{\lambda }}}^{\ast }\left( \theta \right) \varphi
_{^{l_{\rho }^{\prime }l_{\lambda }^{\prime }}}\left( \theta \right) +\xi
\left( I_{a}^{+}+I_{a}^{-}\right) \right]  \label{(2.16)}
\end{equation}

\bigskip

\bigskip With:

\begin{eqnarray}
I_{b}^{\pm } &=&\dint d\Omega _{5}\left\vert \overrightarrow{r_{1}}\pm 
\overrightarrow{r_{2}}\right\vert ^{-1}Y_{l_{\rho }l_{\lambda }}^{M\ast
}\left( \Omega _{5}\right) Y_{l_{\rho }^{\prime }l_{\lambda }^{\prime
}}^{M^{\prime }}\left( \Omega _{5}\right)  \label{(2.17)} \\
&=&\dint\limits_{0}^{\frac{\pi }{2}}d\theta \sin ^{2}\theta \cos ^{2}\theta
\varphi _{^{l_{\rho }l_{\lambda }}}^{\ast }\left( \theta \right) \varphi
_{^{l_{\rho }^{\prime }l_{\lambda }^{\prime }}}\left( \theta \right) \dint
d\Omega _{\rho }Y_{l_{\rho }}^{m_{\rho }\ast }\left( \Omega _{\rho }\right)
Y_{l_{\rho }^{\prime }}^{m_{\rho }^{\prime }}\left( \Omega _{\rho }\right) 
\notag \\
&&\times \dint d\Omega _{\lambda }Y_{l_{\lambda }}^{m_{\lambda }\ast }\left(
\Omega _{\lambda }\right) Y_{l_{\lambda }^{\prime }}^{m_{\lambda }^{\prime
}}\left( \Omega _{\lambda }\right) \dsum\limits_{l=0}^{\infty
}\dsum\limits_{m=-l}^{\infty }\left( \mp \right) ^{l}b_{l}\frac{4\pi }{2l+1}%
Y_{l}^{m\ast }\left( \Omega _{\rho }\right) Y_{l}^{m}\left( \Omega _{\lambda
}\right)  \notag
\end{eqnarray}

\begin{eqnarray}
I_{a}^{\pm } &=&\dint d\Omega _{5}\left\vert \overrightarrow{r_{1}}\pm 
\overrightarrow{r_{2}}\right\vert Y_{L}^{M\ast }\left( \Omega _{5}\right)
Y_{L^{\prime }}^{M^{\prime }}\left( \Omega _{5}\right)  \label{(2.18)} \\
&=&\dint\limits_{0}^{\frac{\pi }{2}}d\theta \sin ^{2}\theta \cos ^{2}\theta
\varphi _{^{l_{\rho }l_{\lambda }}}^{\ast }\left( \theta \right) \varphi
_{^{l_{\rho }^{\prime }l_{\lambda }^{\prime }}}\left( \theta \right) \dint
d\Omega _{\rho }Y_{l_{\rho }}^{m_{\rho }\ast }\left( \Omega _{\rho }\right)
Y_{l_{\rho }^{\prime }}^{m_{\rho }^{\prime }}\left( \Omega _{\rho }\right) 
\notag \\
&&\times \dint d\Omega _{\lambda }Y_{l_{\lambda }}^{m_{\lambda }\ast }\left(
\Omega _{\lambda }\right) Y_{l_{\lambda }^{\prime }}^{m_{\lambda }^{\prime
}}\left( \Omega _{\lambda }\right) \dsum\limits_{l=0}^{\infty
}\dsum\limits_{m=-l}^{\infty }\left( \mp \right) ^{l}a_{l}\frac{4\pi }{2l+1}%
Y_{l}^{m\ast }\left( \Omega _{\rho }\right) Y_{l}^{m}\left( \Omega _{\lambda
}\right)  \notag
\end{eqnarray}

\bigskip

Where :

\begin{equation}
\varphi _{^{l_{\rho }l_{\lambda }}}\left( \theta \right) =D_{^{l_{\rho
}l_{\lambda }}}\sin ^{l_{\rho }}\theta \sin ^{l_{\lambda }}\theta
\label{(2.19)}
\end{equation}

$\bigskip D$ $_{L}^{l_{\rho }l_{\lambda }}$ : is the normalization factor.

\ \ 
\begin{equation}
\ b_{l}=\frac{r_{<}^{l}}{r_{>}^{l+1}}\ ,\ \ a_{l}=\frac{1}{2l+3}\frac{%
r_{<}^{l+2}}{r_{>}^{l+1}}-\frac{1}{2l-1}\frac{r_{<}^{l}}{r_{>}^{l-1}}
\label{(2.20)}
\end{equation}

\ \ $(\ r_{<}=\min (r_{1},r_{2})$ , $r_{>}=\max (r_{1},r_{2})$ $)$ .\bigskip

From our choice of potential ( Coulombian + Linear ) we have :

\begin{equation}
\overrightarrow{r_{1}}=\frac{\sqrt{3}}{\sqrt{2}}\overrightarrow{\lambda }%
\text{ , }\overrightarrow{r_{2}}=\frac{1}{\sqrt{2}}\overrightarrow{\rho }
\label{(2.21)}
\end{equation}

Expanding the two parts $I_{b}^{\pm }$ and $I_{a}^{\pm }$, we obtain :

\begin{equation}
I_{b}^{\pm }=\sum_{l=0}^{\min \left[ l_{\rho }+l_{\rho }^{\prime
},l_{\lambda }+l_{\lambda }^{\prime }\right] }\left( \mp \right) ^{l}\left(
-\right) ^{l-m_{\lambda }}\left( -\right) ^{m_{\rho }^{\prime
}}M_{1}M_{2}\times \left[ b_{1}+b_{2}\right]  \label{(2.22)}
\end{equation}

\begin{equation}
I_{a}^{\pm }=\sum_{l=0}^{\min \left[ l_{\rho }+l_{\rho }^{\prime
},l_{\lambda }+l_{\lambda }^{\prime }\right] }\left( \mp \right) ^{l}\left(
-\right) ^{l-m_{\lambda }}\left( -\right) ^{m_{\rho }^{\prime
}}M_{1}M_{2}\times \left[ a_{1}+a_{2}\right]  \label{(2.23)}
\end{equation}

With :

\begin{equation}
M_{1}=i^{l_{\rho }^{\prime }+l-l_{\rho }}\sqrt{\left( 2l_{\rho }+1\right)
\left( 2l_{\rho }^{\prime }+1\right) }\left( 
\begin{array}{ccc}
l_{\rho } & l_{\rho }^{\prime } & l \\ 
-m_{\rho } & 0 & m_{\rho }%
\end{array}%
\right) \left( 
\begin{array}{ccc}
l_{\rho } & l_{\rho }^{\prime } & l \\ 
0 & 0 & 0%
\end{array}%
\right)  \label{(2.24)}
\end{equation}

\begin{equation}
M_{2}=i^{l_{_{\lambda }}^{\prime }+l-l_{_{\lambda }}}\sqrt{\left(
2l_{\lambda }+1\right) \left( 2l_{\lambda }^{\prime }+1\right) }\left( 
\begin{array}{ccc}
l_{\lambda } & l_{\lambda }^{\prime } & l \\ 
-m_{\lambda } & 0 & m_{\lambda }%
\end{array}%
\right) \left( 
\begin{array}{ccc}
l_{\lambda } & l_{\lambda }^{\prime } & l \\ 
0 & 0 & 0%
\end{array}%
\right)  \label{(2.25)}
\end{equation}

\begin{equation}
b_{1}=\dint\limits_{0}^{\arctan \left( 2\right) }d\theta \sin ^{2}\theta
\cos ^{2}\theta \Psi _{L}^{l_{\rho }l_{\lambda }\ast }\left( \theta \right)
\Psi _{L^{\prime }}^{l_{\rho }^{\prime }l_{\lambda }^{\prime }\ast }\left(
\theta \right) \times \left( \frac{2}{3^{l+1}}\right) ^{\frac{1}{2}}\left( 
\frac{\sin ^{l}\theta }{\cos ^{l+1}\theta }\right)  \label{(2.26)}
\end{equation}

\begin{equation}
b_{2}=\dint\limits_{\arctan \left( 2\right) }^{\frac{\pi }{2}}d\theta \sin
^{2}\theta \cos ^{2}\theta \Psi _{L}^{l_{\rho }l_{\lambda }\ast }\left(
\theta \right) \Psi _{L^{\prime }}^{l_{\rho }^{\prime }l_{\lambda }^{\prime
}\ast }\left( \theta \right) \times \left( 2\times 3^{l}\right) ^{\frac{1}{2}%
}\left( \frac{\cos ^{l}\theta }{\sin ^{l+1}\theta }\right)  \label{(2.27)}
\end{equation}

\begin{eqnarray}
a_{1} &=&\dint\limits_{0}^{\arctan \left( 2\right) }d\theta \sin ^{2}\theta
\cos ^{2}\theta \Psi _{L}^{l_{\rho }l_{\lambda }\ast }\left( \theta \right)
\Psi _{L^{\prime }}^{l_{\rho }^{\prime }l_{\lambda }^{\prime }\ast }\left(
\theta \right)  \label{(2.28)} \\
&&\times \left[ \left( \frac{1}{2\times 3^{l+1}}\right) ^{\frac{1}{2}}\frac{%
\sin ^{l+2}\theta }{\left( 2l+3\right) \cos ^{l+1}\theta }-\left( \frac{1}{%
2\times 3^{l-1}}\right) ^{\frac{1}{2}}\frac{\sin ^{l}\theta }{\left(
2l-1\right) \cos ^{l-1}\theta }\right]  \notag
\end{eqnarray}

\begin{eqnarray}
a_{2} &=&\dint\limits_{\arctan \left( 2\right) }^{\frac{\pi }{2}}d\theta
\sin ^{2}\theta \cos ^{2}\theta \Psi _{L}^{l_{\rho }l_{\lambda }\ast }\left(
\theta \right) \Psi _{L^{\prime }}^{l_{\rho }^{\prime }l_{\lambda }^{\prime
}\ast }\left( \theta \right)  \label{(2.29)} \\
&&\times \left[ \left( \frac{3^{l+2}}{2}\right) ^{\frac{1}{2}}\frac{\cos
^{l+2}\theta }{\left( 2l+3\right) \sin ^{l+1}\theta }-\left( \frac{3^{l}}{2}%
\right) ^{\frac{1}{2}}\frac{2^{l-1}\cos ^{l}\theta }{\left( 2l-1\right) \sin
^{l-1}\theta }\right]  \notag
\end{eqnarray}

\bigskip

Finally the total average potential energy is the sum of matrix elements $%
V_{l_{\rho }l_{\lambda },l_{\rho }^{\prime }l_{\lambda }^{\prime}}$
multiplied by Clebsch-Gordan coefficients :

\begin{equation}
V_{LM,L^{\prime}M^{\prime}}=\underset{m_{\rho },m_{\lambda }}{\underset{%
l_{\rho },l_{\lambda }}{\dsum }}\underset{m_{\rho }^{\prime },m_{\lambda
}^{\prime }}{\underset{l_{\rho }^{\prime },l_{\lambda }^{\prime }}{\dsum }}%
\left\langle l_{\rho }m_{\rho }l_{\lambda }m_{\lambda }|LM\right\rangle
\left\langle l_{\rho }^{\prime }m_{\rho }^{\prime }l_{\lambda }^{\prime
}m_{\lambda }^{\prime }|L^{\prime }M^{\prime }\right\rangle \times
V_{l_{\rho }l_{\lambda },l_{\rho }^{\prime }l_{\lambda }^{\prime }}
\label{(2.30)}
\end{equation}

With : 
\begin{eqnarray}
\left\{ 
\begin{array}{r@{\quad ,\quad}l}
\overrightarrow{L}=\overrightarrow{l_{\rho }}+\overrightarrow{l_{\lambda }}
& \ \ (M=m_\rho +m_\lambda) \\ 
\overrightarrow{L^{\prime }}=\overrightarrow{l_{\rho}^{\prime }}+%
\overrightarrow{l_{\lambda }^{\prime }} & \ \ (M^{\prime }=m_{\rho}^{\prime
} +m_{\lambda}^{\prime })%
\end{array}
\right.  \notag \\
\notag
\end{eqnarray}

The matrix elements $V_{l_{\rho }l_{\lambda },l_{\rho }^{\prime }l_{\lambda
}^{\prime}}$ lead to enormous calculus, so we are in the obligation to
restrict ourselves to the lower orbital excitations (pratically we limit
ourselves to $(l_{\rho },l_{\lambda },l_{\rho }^{\prime },l_{\lambda
}^{\prime }\leq 2)$ ). Doing such restriction leads to a very simple
analytically expression wich can be solved.

\ 

\subsection{Some relations}

\begin{equation}
Y_{l}^{m\ast }\left( \Omega \right) =\left( -1\right) ^{l-m}Y_{l}^{(-m)\ast
}\left( \Omega \right)  \label{(2.31)}
\end{equation}

\begin{equation}
\int d\Omega Y_{l}^{m\ast }\left( \Omega \right) Y_{l^{\prime }}^{m^{\prime
}}\left( \Omega \right) =\delta _{ll^{\prime }}\delta _{mm^{\prime }}
\label{(2.32)}
\end{equation}

\begin{eqnarray}
\int d\Omega Y_{l_{1}}^{m_{1}\ast }\left( \Omega \right) Y_{l}^{m}\left(
\Omega \right) Y_{l_{2}}^{m_{2}\ast } &=&\left( -1\right)
^{m}i^{l+l_{2}-l_{1}} \\
&&\times \sqrt{\frac{\left( 2l_{1}+1\right) \left( 2l+1\right) \left(
2l_{2}+1\right) }{4\pi }}  \notag \\
&&\times \left( 
\begin{array}{ccc}
l_{1} & l & l_{2} \\ 
-m_{1} & 0 & m_{2}%
\end{array}%
\right) \left( 
\begin{array}{ccc}
l_{1} & l & l_{2} \\ 
0 & 0 & 0%
\end{array}%
\right)  \notag
\end{eqnarray}

\begin{equation}
\sum_{m_{1}m_{2}}\left( 
\begin{array}{ccc}
l_{1} & l_{2} & L \\ 
m_{1} & m_{2} & -M%
\end{array}%
\right) \left( 
\begin{array}{ccc}
l_{1} & l_{2} & L^{\prime } \\ 
m_{1} & m_{2} & -M^{\prime }%
\end{array}%
\right) =\frac{1}{2L+1}\delta _{LL^{\prime }}\delta _{MM^{\prime }}
\end{equation}


\end{document}